\documentclass[aps,pra,preprint,letter,showpacs,floatfix]{revtex4}
\usepackage{graphicx,amsmath,amssymb,textcomp}
\renewcommand{\d}{\mathrm{d}}
\newcommand{\e}{\mathrm{e}}
\renewcommand{\Re}{\mathrm{Re}}

\begin{document}
\title{Four-photon scattering in birefringent fibers}
\author{E. Brainis}
\email{ebrainis@ulb.ac.be}
\affiliation{Department of Physics, University of Oxford, Clarendon
Laboratory, Parks Road OX1 3PU, Oxford, United Kingdom}
\begin{abstract}
Four-photon scattering in nonlinear waveguides is an important
physical process that allows photon-pair generation in well defined
guided modes, with high rate and reasonably low noise. Most of the experiments
to date used the scalar four-photon scattering process in which the
pump photons and the scattered photons have the same polarization.
In birefringent waveguides, vectorial four-photon scattering are
also allowed: these vectorial scattering processes involve photons
with different polarizations. In this article, the theory of
four-photon scattering in nonlinear, birefringent, and dispersive
fibers is developed in the framework of the quantum theory of light.
The work focusses on the spectral properties and quantum correlations (including entanglement) of
photon-pairs generated in high-birefringence and low-birefringence fibers.
\end{abstract}
 \maketitle
\section{Introduction}
When two intense monochromatic pump beams (or quasi-monochromatic pulses) are
 launched together in an one-dimensional nonlinear $\chi^{(3)}$ medium,
 matter-light interaction results in the scattering of pump
photons to other wavelengths. The main scattering
 process consists in the spontaneous conversion of two pump photons
with angular frequencies $\omega_{01}$ and $\omega_{02}$ into a
red-shifted Stokes photon (signal) and a blue-shifted anti-Stokes
photon (idler) with angular frequencies $\omega_s$ and $\omega_a$,
satisfying $\omega_{01}+\omega_{02}=\omega_s+\omega_a$.
 This elastic process in known as a \emph{four-photon scattering} (FPS). It conserves the number
of particles as well as the total energy in the field.

During the recent years, FPS has attracted much attention from experimentalists because it
allows to generate correlated, and sometimes entangled, photon pairs
in optical fibers \cite{fiorentino2002, Li2004,Sharping2004,
Takesue2005, Rarity2005, Li2005, FanDog2005, Fan2005, Fan2005b,
Fulconis2005,chen2005, Lee2006, alibart2006, Takesue2006, Chen2006,
Fan2007a, Fan2007b, Fan2007c, Li2008a, Li2008b, Dyer2008}. These fiber-optics photon-pair sources constitute a major
progress in the development of quantum photonics because they offer the
advantage of being compact guided-wave sources that can be easily
connected (with only marginal losses) to standard transmission
fibers. This constitutes an important advantage for applications 
in fiber quantum
communication (quantum key distribution \cite{Gisin}, remote coin
tossing \cite{Lamoureux2005}) and fiber quantum computation
\cite{Brainis2003}. Recently, FPS in silicon waveguides
\cite{Sharping2006,Takesue2007, Takesue2008} has also been
demonstrated and used to produce photon-pairs suitable for
silicon-on-insulator quantum circuits.

In contrast with the considerable amount of experiments,
theoretical work on FPS has been rather limited \cite{Wang2001,chen2005,
Lin2006,Lin2007}. The early work of
Wang et al. \cite{Wang2001} gives a quantum description of FPS in
the \emph{scalar} case, where the pump photons and the photon pairs
share the same polarization. That treatment is restricted to a
single monochromatic pump (degenerate case,
$\omega_{01}=\omega_{02}$) and was generalized later \cite{chen2005} to 
account for the spectral shape of
short pump pulses. In addition, Ref.~\onlinecite{chen2005} introduced, 
for the first time, first-order perturbation theory for studying FPS. 
Perturbation theory turns out to be a very suitable approach
 and will be used extensivelly in this work.
 In 2006, Lin et al.
\cite{Lin2006,Lin2007}  examined the theory of Raman noise in
fiber-optics photon-pair sources and observed that the impact of
Raman noise can be reduced by producing photon pairs in birefringent 
fibers through a
\emph{vectorial} FPS which
generates photons that are orthogonally polarized with respect to the pump.
 So far,
vectorial FPS processes have not been investigated experimentally (except in
\cite{Fan2005b} in a reverse degenerate FPS configuration).

In this paper, the theory of FPS in a nonlinear, birefringent, and
dispersive fiber is developed  in the framework of the quantum
theory of light. The quantum state of light at the output of the fiber in computed using first-order perturbation theory.  
Only degenerate FPS from a single monochromatic 
pump ($\omega_{01}=\omega_{02}\equiv\omega_0$) is considered and 
 Raman scattering is not included. 
The work focus on the \emph{spectral properties} of
the generated photon-pairs and the kind of \emph{correlations} that can
be obtained in vectorial FPS processes.

If the optical power of the pump and the propagation length are
large enough, the photons that have been generated by FPS may be
amplified by stimulated four-wave mixing (FWM). FPS
and stimulated FWM are two different aspects of the same phenomenon.
 To highlight this point, the properties of vectorial FPS
are derived from the
quantum nonlinear Schr\"odinger equations
(see Sec.~\ref{sec:3}), in close analogy with the classical analysis of 
modulation instabilities (self phase-mached FWM processes in fibers). 
The cases of high-birefringence and
low-birefringence fibers are treated separately (Sec.~\ref{sec:4}
and Sec.~\ref{sec:5}), as in the classical theory \cite{agraw}. 
Although, FPS and classical FWM are strongly related,
they are still physically distinct processes because, as will be
shown in Sec.~\ref{sec:6}, parametric gain is not required for FPS.
Therefore, the spectrum of photons generated by FPS is usually wider
than the range of frequencies experiencing parametric amplification
due to classical FWM.

\section{Quantum theory of nonlinear propagation}\label{sec:2}
Single mode optical fibers are one-dimensional propagation media
that are weakly dispersive, weakly nonlinear and possibly
birefringent. Only \emph{linear} birefringence is considered here.
The propagating electric field operator (in the Heisenberg picture)
can be written as
\begin{equation}
\mathbf{E}=\sum_{j\in\{x,y\}} E_j(z,t)\ \mathbf{e}_j \
\mathcal{F}(x,y),
\end{equation}
where $\mathcal{F}(x,y)$ ($\iint |\mathcal{F}(x,y)|^2 \d x \d y=1$)
is a function of the transverse coordinates describing the profile
of the guided mode (it is assumed to be the same for both
polarizations and all the frequencies involved), $\mathbf{e}_j$
($j\in\{x,y\}$) are unit vectors along the optical axes, and
\begin{equation}\label{E1}
E_j(z,t)=\int_0^\infty \sqrt{\frac{\hbar \omega}{2\epsilon_0
n_j(\omega) c}} \ b_j(z,\omega) \ \frac{\e^{i(k_{j}(\omega)z-\omega
t)}}{\sqrt{2\pi}} \ \d\omega.
\end{equation}
Here $k_{j}(\omega)$ and $n_j(\omega)$ are the propagation constant
and refraction index on the $j$-axis. The operators $b_j(z,\omega)$
are ordinary photon annihilation operators. In absence of
any non-linearity, the operators $b_j(z,\omega)$ are $z$-independent. However, the
propagation of a strong monochromatic pump in the fiber (with angular frequency $\omega_{0}$) makes the different $b_j(z,\omega)$
$z$-dependant. For any $\omega\neq\omega_0$, the $z$-dependance can
be split in two parts: (i) a FWM part that, in the quantum theory, also 
 gives rise to
the FPS phenomenon and (ii) a nonlinear phase shift due to the 
cross phase modulation (CPM) induced by the pump. To
make the CPM effect explicit, we write
$b_j(z,\omega)=b'_j(z,\omega)\exp{[i \ \phi_{\mathrm{CPM}}(z)]}$,
where
\begin{equation}\label{cpm}
\phi_{\mathrm{CPM}}(z)= 2 \  \gamma
\left(P_{0j}+\frac{1}{3}P_{0\bar{j}}\right).
\end{equation}
\emph{The $z$-evolution of $b'_j(z,\omega)$ is only due to
wave-mixing}. The phase factor due to the CPM depends on the
polarization of the pump. $P_{0j}$ and $P_{0\bar{j}}$ are the pump
powers propagating on the $j$-axis and the $\bar{j}$-axis,
respectively ($\bar{j}=y$ if $j=x$ and vice versa). The factor
$\gamma$ is the non-linearity parameter of the fiber, defined in
Sec.~\ref{sec:3} [Eq.~(\ref{gamma})].

Because the fields that will be considered have a small bandwidth
(usually up to several tens of THz) around the central pump
frequency $\omega_0$, one can neglect the frequency dependence of
the square root in the right-hand side of Eq.~(\ref{E1}) and write
\begin{equation}\label{modal1}
E_j(z,t)=\sqrt{\frac{\hbar \omega_0}{2\epsilon_0 n_{j0} c}}
\  \e^{-i\omega_0 t}  \int_{-\infty}^\infty  a_j(z,\Omega) \ \frac{\e^{i(\beta_{j}(\Omega)z+
\phi_{\mathrm{CPM}}(z))-i \Omega t}}{\sqrt{2\pi}} \d\Omega,
\end{equation}
where $n_{j0}=n_j(\omega_0)$ and $\Omega=\omega-\omega_0$ is the
frequency detuning. We also set $a_j(z,\Omega)\equiv
b'_j(z,\omega)$ and $\beta_{j}(\Omega)\equiv k_j(\omega)$. The
operators $a_j(z,\Omega)$ and $a_j^\dag(z,\Omega)$ satisfy the
usual bosonic commutation relations:
\begin{subequations}\label{commutbos}
\begin{eqnarray}
\left[a_j^{\phantom{\dag}}(z,\Omega),a^\dag_{j'}(z,\Omega')\right]&=& \delta_{jj'} \ \delta(\Omega-\Omega'),\\
\left[a_j^{\phantom{\dag}}(z,\Omega),a_{j'}^{\phantom{\dag}}(z,\Omega')\right]&=&0,\\
\left[a_j^{\dag}(z,\Omega),a^\dag_{j'}(z,\Omega')\right]&=&0.
\end{eqnarray}
\end{subequations}
 Note that the fiber is not infinite: it begins
at $z=0$ and end at $z=L$. Since the linear dispersion relations are
discontinuous at these two points, a mode-coupling between forward
and backward propagating photons occurs (Fresnel reflection due to
impedance mismatch). We will neglect these effects and consider that
$a_j(0,\Omega)=\lim_{z\rightarrow -\infty}a_j(z,\Omega)$ and $a_j(L,\Omega)=\lim_{z\rightarrow
\infty}a_j(z,\Omega)$ (there is no non-linearity outside the fiber).

In order to solve the nonlinear propagation problem, we need
the unitary operator $U(L,0)$ that maps
$a_j(0,\Omega)$ on
$a_j(L,\Omega)$ for any detuning $\Omega$:
\begin{equation}\label{transf}
a_j(L,\Omega)=U^\dag(L,0)a_j(0,\Omega)U(L,0).
\end{equation}
Obviously, the evolution of
$a_j$ from $z=0$ to $z=L$ is a continuous process.
Therefore, there exists an unitary \emph{evolution operator}
$U(z_2,z_1)$ such that
\begin{equation}\label{cont}
a_j(z_2,\Omega)=U^\dag(z_2,z_1)a_j(z_1,\Omega)U(z_2,z_1)
\end{equation}
for any couple of coordinates $(z_1,z_2)$ in $[0,L]$. The continuity
in the variables $z_1$ and $z_2$ implies the existence of an
Hermitian operator $G(z)$ (the ``infinitesimal generator'') such
that, for small $\delta z$, $U(z+\delta z,z)= 1 + \frac{i}{\hbar}
G(z) \delta z+o(\delta z)$. Using this last relation and
Eq.~(\ref{cont}), one finds that the annihilation operators
$a_j(z,\Omega)$ satisfies the following \emph{Heisenberg equations}:
\begin{equation}\label{heisen}
i\hbar \frac{\partial}{\partial z}
a_j(z,\Omega)=\left[G(z),a_j(z,\Omega)\right].
\end{equation}
As it will be shown later, Eqs.~(\ref{heisen}) are just the usual
coupled-mode equations of nonlinear optics. In principle, solving
these equations yields the solutions $a_j(L,\Omega)$. The operators
$a_j(L,\Omega)$ and the initial quantum state of light
$|\psi\rangle$ provide a complete knowledge about the final state
since the mean number of photons $\langle\psi|a_j^\dag(L,\Omega)
a_j(L,\Omega)|\psi\rangle$ in each mode can be computed, as well as
any photon correlation between modes.

An alternative approach consists in working in the \emph{interaction
picture} instead of the usual Heisenberg picture. The passage from
one picture to the other one is performed according to the transformation
\begin{subequations}\label{IH}
\begin{gather}
|\psi^I(z)\rangle=U(z,0)|\psi\rangle,\\
\ A^I(z)=U(z,0)A(z)U^\dag(z,0)
\end{gather}
\end{subequations}
where $|\psi\rangle$ is an arbitrary initial quantum state of light and $A$
an arbitrary quantum operator in the Heisenberg picture. For
instance, it follows from Eq.~(\ref{cont}) that \emph{the
interaction-picture annihilation operators are $z$-independant}:
\begin{equation}
a_j^I(z,\Omega)=U(z,0)a_j(z,\Omega)U^\dag(z,0)=a_j(0,\Omega)\equiv
a_j(\Omega).
\end{equation}
As a general rule, the interaction-picture evolution
of operators is due to dispersion, birefringence (the linear
properties of the fiber) and CPM (but not FWM), while the interaction-picture 
evolution of the quantum
states is only due to the FWM:
\begin{equation}\label{S}
-i\hbar \frac{\d}{dz}|\psi^I(z)\rangle= G^I(z)|\psi^I(z)\rangle,
\end{equation}
where
\begin{equation}\label{GI}
G^I(z)=U(z,0)G(z)U^\dag(z,0).
\end{equation}
From (\ref{IH}a), we see that $|\psi\rangle=|\psi^I(z=0)\rangle$.
Inserting (\ref{IH}a) into (\ref{S}) gives an equation for $U(z,0)$:
\begin{equation*}
\frac{\partial}{\partial z}U(z,0)=\frac{i}{\hbar} \ G^I(z) \ U(z,0),
\end{equation*}
which can also be written as an integral equation:
\begin{equation}\label{Uimpl}
U(L,0)=1+\frac{i}{\hbar}\int_0^L \ G^I(z') \ U(z',0) \ \d z'.
\end{equation}
The main advantage of working in the interaction picture is that
Eq.~(\ref{Uimpl}) can be solved iteratively using the standard
Dyson's perturbation expansion \cite{joachainv2} from quantum
theory of scattering. For vanishing non-linearity,
the second term of the right-hand side of (\ref{Uimpl}) can be neglected, 
so the zero-order approximation of $U(L,0)$ is just
$U^{(0)}(L,0)=1$. For weak non-linearity, a better approximation 
is obtained by replacing
$U(z',0)$ under the integral sign in Eq.~(\ref{Uimpl}) by the
zero-order approximation $U^{(0)}(z',0)=1$. This yields the
first-order approximation
\begin{equation}\label{first}
U^{(1)}(L,0)=1+\frac{i}{\hbar}\int_0^L \ G^I(z')  \ \d z'.
\end{equation}
Continuing the recurrence, better approximations can be found using:
\begin{equation}\label{norder}
U^{(n)}(L,0)=1+\frac{i}{\hbar}\int_0^L \ G^I(z') \ U^{(n-1)}(z',0) \
\d z'.
\end{equation}

Since a single coherent monochromatic pump field is launched in the
optical fiber, the initial state $|\psi^I(z=0)\rangle$ is the vacuum
state for all modes \emph{except the pump mode}. In the article, we
call this state $|0\rangle$. As we will show later, using
Eq.~(\ref{IH}a) with the first-order approximation of the evolution
operator $U^{(1)}(L,0)$ yields the final state
\begin{eqnarray}\label{FFS}
|\psi^I(L)\rangle=|0\rangle&+&\int_0^\infty \xi_{xx}(L,\Omega) \
|1^x_\Omega,1^x_{-\Omega}\rangle  \  \d\Omega
\nonumber \\  
&+&\int_0^\infty \xi_{yy}(L,\Omega) \
|1^y_\Omega,1^y_{-\Omega}\rangle \ \d\Omega \nonumber \\
&+&\int_0^\infty \xi_{xy}(L,\Omega) \
|1^x_\Omega,1^y_{-\Omega}\rangle \ \d\Omega \nonumber \\
&+&\int_0^\infty \xi_{yx}(L,\Omega) \
|1^y_\Omega,1^{x}_{-\Omega}\rangle \ \d\Omega,
\end{eqnarray}
where
$|1^j_\Omega,1^{j'}_{-\Omega}\rangle=a_j^\dag(\Omega)a_{j'}^\dag(-\Omega)|0\rangle$.
Two photons are created as a consequence of the annihilation of two
pump photons. This process is called a FPS. The
\emph{two-photon amplitudes} $\xi_{jj'}(z,\Omega)$ determine the
polarization and spectral properties of the created photon pair.

Since a continuous-mode formalism is used, the states
$|1^j_\Omega,1^{j'}_{-\Omega}\rangle$ are not dimensionless (they
have the dimension of time) and cannot be normalized in the usual
sense. (Only linear combinations of a continuous set of these states
are physical.) When the pump has a finite duration $T$ (but is still
long enough to allow the monochromatic-pump approximation), it
is often easier to describe the scattering in terms of discrete
modes, the spectral width of which is equal to $\Delta\omega=2\pi/T$. The
transition from the continuous-mode to the discrete-mode description
requires the substitutions
\begin{gather*}
\int \ ... \ \d\Omega\rightarrow \sum_{\Omega} \ ... \
\Delta\omega\\
a_j(\Omega) \ \sqrt{\Delta\omega} \rightarrow a_j(\Omega).
\end{gather*}
In the discrete-mode description the two-photon states
$|1^j_\Omega,1^{j'}_{-\Omega}\rangle$ are dimensionless and
normalized.

It is important to note that the whole perturbation scheme
 strongly depends on the way the
operators $a_j(L,\Omega)$ are defined by Eq.~(\ref{modal1}). In
particular, \emph{the fact that we factored out the CPM phase factor} makes
the perturbation scheme used in this work different from the one
used in the previous perturbation-theory description of FPS
\cite{chen2005}. The perturbation scheme used
in this work gives more precise results than the approach
in \cite{chen2005} at any perturbation order \cite{Brainis2008}.

Using the theory presented in the next sections, we will be able to
compute the two-photon amplitudes $\xi_{jj'}(z,\Omega)$ and show the
connection between FPS and classical FWM.

\section{The quantum nonlinear Schr\"odinger equations}\label{sec:3}
If the non-linearity is neglected, the fiber exhibits two linearly
polarized optical modes with propagations constants $\beta_x$ and
$\beta_y$.
Because of the dispersion, both propagation constants vary slowly
with the angular frequency of light. In many cases, a second order
Taylor approximation around the pump frequency $\omega_0$ is precise
enough to capture the essence of the physics involved. We can write
\begin{equation}\label{disp}
\beta_{j}(\Omega)=\beta_{0j}+\beta_{1j}\Omega+\frac{\beta_{2j}}{2}\Omega^2,
\end{equation}
with $\beta_{nj}=\d^n\beta_j/\d\Omega^n(\Omega=0)$ and $j\in\{x,y\}$.
The two polarization modes have thus different phase and group
velocities. The parameters $\Delta\beta_0=\beta_{0x}-\beta_{0y}$ and
$\Delta\beta_1=\beta_{1x}-\beta_{1y}$ measure the phase and group
velocity mismatch, respectively. We will assume that the
group-velocity dispersion (GVD) parameter [$\beta_{2j}$ 
in Eq.~(\ref{disp})] is the
same for both axes and write $\beta_2\equiv \beta_{2x}=\beta_{2y}$. 
 When $2\pi c/\omega_0$ is close to the so-called zero-dispersion wavelength,
$\beta_2$ can be so small that higher-order terms in the Taylor
expansion (\ref{disp}) must to be taken into account for
consistency. Although this situation is encountered in many
experiments that generate far detuned photon pairs using scalar FPS,
we will not consider that case in this study of vector FPS.

The intrinsic non-linearity of silica fibres is of third order and
can be considered isotropic \cite{agraw}. As a consequence, the
nonlinear susceptibility tensor has only one independent element
$\chi_{xxxx}$. The non-linearity parameter of the fibre is defined
by:
\begin{equation}\label{gamma}
\gamma=\frac{3\omega_0\chi^{(3)}_{xxxx}}{4\epsilon_0
n^2c^2A_{\mathrm{eff}}},
\end{equation}
where $n$ and $A_{\mathrm{eff}}$ are, respectively, the effective
linear index and the effective area of the fibre at the frequency
$\omega_0$. The constants $\epsilon_0$ and $c$ are the vacuum 
permittivity and velocity of light.

Light propagation in a single-mode
birefringent fiber is properly described by the a set of coupled
nonlinear Schr\"{o}dinger equations \cite{agraw} for the
\emph{slowly varying envelopes} $A_x$ and $A_y$ of the two
polarization modes.  
These are connected to the electric field
components by the relations
\begin{equation}\label{A}
E_j(z,t)=\mathcal{N} \ A_j(z,t)\ \e^{i(\beta_{0j}z-\omega_0 t)}.
\end{equation}
Fast time oscillation at the pump frequency $\omega_0$ and fast
space oscillations with the pump propagation constants $\beta_{0x}$
and $\beta_{0y}$ have been factored out. In quantum theory,
the envelope
fields are operators. The dimensional constant
$\mathcal{N}=1/\sqrt{2\epsilon_0 n_{j0}c}$ is chosen so that
$\langle A_j^\dag(z_0,t)A_j^{\phantom{\dag}}(z_0,t)\rangle$ is the
mean optical power flowing through the plane $z=z_0$ at time $t$
with a polarization along the $j$-axis, $j\in\{x,y\}$. If the approximations explained above hold, the coupled
quantum nonlinear Schr\"{o}dinger equations are
\begin{widetext}
\begin{subequations}\label{QNLS}
\begin{eqnarray}
\frac{\partial A_x}{\partial z}+\beta_{1x} \ \frac{\partial
A_x}{\partial t}+i \ \frac{\beta_2}{2} \ \frac{\partial^2
A_x}{\partial t^2}&=& i \ \gamma \ \left(A_x^\dag
A_x^{\phantom{\dag}}+\frac{2}{3} \ A_y^\dag
A_y^{\phantom{\dag}}\right)A_x+
i \ \frac{\gamma}{3} \ A_y^2A_x^\dag \  \e^{-2i\Delta\beta_0z},\\
\frac{\partial A_y}{\partial z}+\beta_{1y} \ \frac{\partial
A_y}{\partial t}+i \ \frac{\beta_2}{2} \ \frac{\partial^2
A_y}{\partial t^2}&=&i \ \gamma \ \left(A_y^\dag
A_y^{\phantom{\dag}}+\frac{2}{3} \ A_x^\dag
A_x^{\phantom{\dag}}\right)A_y+i \ \frac{\gamma}{3} \ A_x^2A_y^\dag
\ \e^{+2i\Delta\beta_0z}.
\end{eqnarray}
\end{subequations}
\end{widetext}

In the following, we want to describe how photons are scattered from
an intense monochromatic pump wave. The pump wave must be a known
steady-state solution of Eq.~(\ref{QNLS}). In principle, there is no
need for the pump being polarized along an optical axis. Its
polarization state can rotate. The general expression of the
steady-state solution of (\ref{QNLS}) is known \cite{agraw}.
However, it involves elliptic functions and is difficult to handel
for the purpose of the present work. We will therefore restrict the
analysis to particular cases of special interest. We will
distinguish high-birefringence (HB) and low-birefringence (LB)
fibers. In the first case, the pump can have any polarization state;
in the second case, we will restrict the analysis to linearly
polarized pump wave parallel to an optical axis to make the problem
tractable.

\section{High-birefringence fibers}\label{sec:4}
In the HB case, the beat length $2\pi/\Delta\beta_0$ is much smaller
than any other relevant length scale: the phase factors in the last
terms of the right-hand side of Eqs.~(\ref{QNLS}) oscillate so
quickly that they simply average to zero. To a good approximation,
Eqs.~(\ref{QNLS}) can be replaced by
\begin{subequations}\label{HBQNLS}
\begin{eqnarray}
\frac{\partial A_x}{\partial z}&=&-\beta_{1x}\frac{\partial
A_x}{\partial t}-i
\frac{\beta_2}{2} \frac{\partial^2 A_x}{\partial t^2}\nonumber \\
 &\phantom{=}& +i \gamma
\left(A_x^\dag A_x^{\phantom{\dag}}+\frac{2}{3} A_y^\dag
A_y^{\phantom{\dag}}\right)A_x,\\
\frac{\partial A_y}{\partial z}&=&-\beta_{1y}\frac{\partial
A_y}{\partial t}-i
\frac{\beta_2}{2} \frac{\partial^2 A_y}{\partial t^2}\nonumber\\
&\phantom{=}&+i \gamma \left(A_y^\dag
A_y^{\phantom{\dag}}+\frac{2}{3} A_x^\dag
A_x^{\phantom{\dag}}\right)A_y.
\end{eqnarray}
\end{subequations}
An initial classical monochromatic wave has a constant complex
envelope
\begin{subequations}\label{HBini}
\begin{eqnarray}
A_x(0,t)&=&\sqrt{P_{0x}} \ \e^{i\theta_{0x}},\\
A_y(0,t)&=&\sqrt{P_{0y}}  \ \e^{i\theta_{0y}}.
\end{eqnarray}
\end{subequations}
$P_{0x}$ and $P_{0y}$ represent the power of the optical fields
polarized along the $x$- and $y$-axis, respectively.
Eqs.~(\ref{HBQNLS}) admit a classical monochromatic solution that is
compatible with the initial conditions (\ref{HBini}):
\begin{subequations}\label{HBclas}
\begin{eqnarray}
A_x(z,t)&=&\sqrt{P_{0x}} \ \e^{i\theta_{0x}} \ \e^{i\gamma\left( P_{0x}+\frac{2}{3} P_{0y} \right)z},\\
A_y(z,t)&=&\sqrt{P_{0y}} \ \e^{i\theta_{0y}}\ \e^{i\gamma\left(
P_{0y}+\frac{2}{3} P_{0x} \right)z}.
\end{eqnarray}
\end{subequations}
However, this solution is never stable: during the propagation, some
pump photons are destroyed and new photons, at different
frequencies, are created, making the field polychromatic. In order
to demonstrate this point, let's introduce the ansatz
\begin{subequations}\label{HBansatz}
\begin{eqnarray}
A_x(z,t)&=&\left(\sqrt{P_{0x}}+u_x(z,t)\right) \ \e^{i\theta_{0x}} \ \e^{i\gamma\left( P_{0x}+\frac{2}{3} P_{0y} \right)z},\\
A_y(z,t)&=&\left(\sqrt{P_{0y}}+u_y(z,t)\right) \ \e^{i\theta_{0y}} \
\e^{i\gamma\left( P_{0y}+\frac{2}{3} P_{0x} \right)z}
\end{eqnarray}
\end{subequations}
in Eqs.~(\ref{HBQNLS}). The fields $u_j(z,t)$, $j\in\{x,y\}$, can be
seen as perturbations to the stationary solution (\ref{HBclas}). By
injecting the ansatz (\ref{HBansatz}) into Eqs.~(\ref{HBQNLS}) and
retaining only the terms linear in $u_j(z,t)$, we obtain the
following equations:
\begin{subequations}\label{HBpert}
\begin{eqnarray}
\frac{\partial u_x}{\partial z}&=&-\beta_{1x}\frac{\partial
u_x}{\partial t}-i \frac{\beta_2}{2} \frac{\partial^2 u_x}{\partial
t^2}+i\ \gamma \
P_{0x}\left(u_x+u_x^\dag\right)\nonumber \\
&\phantom{=}&+ i\ \frac{2}{3} \ \gamma \ \sqrt{P_{0x}P_{0y}}\left(u_y+u_y^\dag\right),\\
\frac{\partial u_y}{\partial z}&=&-\beta_{1y}\frac{\partial
u_y}{\partial t}-i \frac{\beta_2}{2} \frac{\partial^2 u_y}{\partial
t^2}+i \ \gamma \
P_{0y}\left(u_y+u_y^\dag\right)\nonumber \\
&\phantom{=}&+ i \ \frac{2}{3}\ \gamma \
\sqrt{P_{0x}P_{0y}}\left(u_x+u_x^\dag\right).
\end{eqnarray}
\end{subequations}
\subsection{Coupled-mode equations}
We now make use of Eqs.~(\ref{cpm}), (\ref{modal1}), (\ref{A}), and
(\ref{HBansatz}) to write the mode expansion of the fields $u_j$:
\begin{equation}\label{HBdecomp}
u_j(z,t)=\sqrt{\frac{\hbar\omega_0}{ 2\pi}}
 \ \e^{-i\theta_{0j}} \ \e^{i\gamma
 P_{0j}z} \ \int  \ a_j(z,\Omega) \
\e^{i\left(\beta_j(\omega_0+\Omega)-\beta_{0j}\right)z} \
\e^{-i\Omega t}  \d\Omega.
\end{equation}
Inserting the expansion (\ref{HBdecomp}) into Eqs.~(\ref{HBpert}),
we obtain the coupled-mode equations
\begin{widetext}
\begin{subequations}\label{HBaEvol}
\begin{eqnarray}
\frac{\partial}{\partial z} \ a_x(z,\Omega)&=& i\gamma P_{0x} \
\e^{2i\theta_{0x}} \ \e^{-i(\beta_2\Omega^2+2\gamma P_{0x})z} \
a^\dag_x(z,-\Omega) \nonumber \\
&\phantom{=}&+\ i\frac{2}{3}\gamma\sqrt{P_{0x}P_{0y}}\Big[\
\e^{i(\theta_{0x}-\theta_{0y})} \
\e^{-i[\Delta\beta_1\Omega+\gamma(P_{0x}-P_{0y})] z} \
a_y(z,\Omega)\Big. \nonumber \\ &\phantom{=}&+  \Big. \
\e^{i(\theta_{0x}+\theta_{0y})} \
\e^{-i[\Delta\beta_1\Omega+\beta_2\Omega^2+\gamma (P_{0x}+P_{0y})]
z} \ a^\dag_y(z,-\Omega)\Big],\\
\frac{\partial}{\partial z} \ a_y^{\phantom{\dag}}(z,\Omega)&=& i\gamma P_{0y}\
\e^{2i\theta_{0y}} \ \e^{-i(\beta_2\Omega^2+2\gamma P_{0y})z} \
a^\dag_y(z,-\Omega) \nonumber \\
&\phantom{=}&+ \ i\frac{2}{3}\gamma\sqrt{P_{0x}P_{0y}}\Big[\
\e^{i(\theta_{0y}-\theta_{0x})} \
\e^{-i[-\Delta\beta_1\Omega+\gamma(P_{0y}-P_{0x})] z} \
a_x^{\phantom{\dag}}(z,\Omega)\Big. \nonumber \\
&\phantom{=}&+\Big. \  \e^{i(\theta_{0x}+\theta_{0y})} \
\e^{-i[-\Delta\beta_1\Omega+\beta_2\Omega^2+\gamma(P_{0x}+P_{0y})]
z}\ a^\dag_x(z,-\Omega)\Big].
\end{eqnarray}
\end{subequations}
\end{widetext}
These equations show that the non-linearity couples the $x$- and
$y$-polarized modes of frequency $\omega_0\pm\Omega$. Note that the
evolution of $a_x(z,-\Omega)$ and $a_y(z,-\Omega)$ is obtained from
(\ref{HBaEvol}a) and (\ref{HBaEvol}b) by replacing $\Omega$ by
$-\Omega$. Comparing Eqs.~(\ref{HBaEvol}) with the general formula
(\ref{heisen}) shows that in the HB limit, the infinitesimal
generator is given by
\begin{widetext}
\begin{eqnarray}\label{HBG}
G_{\mathrm{hb}}(z)&=& \frac{\hbar}{2} \gamma P_{0x}\left(\e^{2i\theta_{0x}}\int_{-\infty}^\infty \  \e^{-i(\beta_2\Omega^2+2\gamma P_{0x})z} \ a^\dag_x(z,\Omega)a^\dag_x(z,-\Omega) \ \d\Omega+ \mathrm{H.c.}\right) \nonumber \\
     &+&\frac{\hbar}{2} \gamma P_{0y}\left(\e^{2i\theta_{0y}}\int_{-\infty}^\infty \  \e^{-i(\beta_2\Omega^2+\gamma P_{0y})z} \ a^\dag_y(z,\Omega)a^\dag_y(z,-\Omega) \ \d\Omega+ \mathrm{H.c.}\right) \nonumber \\
     &+& \hbar \frac{2}{3} \gamma \sqrt{P_{0x}P_{0y}} \nonumber \\
     &\phantom{+} &  \left(\e^{i(\theta_{0x}-\theta_{0y})}\int_{-\infty}^\infty \  \e^{-i[\Delta\beta_1\Omega+\gamma(P_{0x}-P_{0y})] z} \ a^\dag_x(z,\Omega)a_y(z,\Omega) \ \d\Omega+ \mathrm{H.c.}\right)\nonumber \\
     &+& \hbar \frac{2}{3} \gamma \sqrt{P_{0x}P_{0y}} \nonumber \\
     &\phantom{+} & \left(\e^{i(\theta_{0y}-\theta_{0x})}\int_{-\infty}^\infty \  \e^{-i[-\Delta\beta_1\Omega+\gamma(P_{0y}-P_{0x})]z} \ a^\dag_y(z,\Omega)a_x(z,\Omega) \ \d\Omega+ \mathrm{H.c.}\right) \nonumber \\
     &+& \frac{\hbar}{3}  \gamma \sqrt{P_{0x}P_{0y}} \nonumber \\
      &\phantom{+} &\left(\e^{i(\theta_{0x}+\theta_{0y})}\int_{-\infty}^\infty \  \e^{-i[\Delta\beta_1\Omega+\beta_2\Omega^2+\gamma (P_{0x}+P_{0y})]z} \ a^\dag_x(z,\Omega)a^\dag_y(z,-\Omega) \ \d\Omega+ \mathrm{H.c.}\right) \nonumber \\
     &+& \frac{\hbar}{3}  \gamma \sqrt{P_{0x}P_{0y}} \nonumber \\
     &\phantom{+} &  \left(\e^{i(\theta_{0x}+\theta_{0y})}\int_{-\infty}^\infty \  \e^{-i[-\Delta\beta_1\Omega+\beta_2\Omega^2+\gamma(P_{0x}+P_{0y})]z} \ a^\dag_y(z,\Omega)a^\dag_x(z,-\Omega) \ \d\Omega+ \mathrm{H.c.}\right).
\end{eqnarray}
\end{widetext}

\subsection{Four-photon
scattering}\label{sec:4:B}
 So far, we used the Heisenberg
picture to derive the coupled-mode equations (\ref{HBaEvol}) and the
corresponding infinitesimal generator (\ref{HBG}). However, as
explained in Sec.~\ref{sec:2}, more physical insight is gained by
working in the \emph{interaction picture}, especially when Dyson's
perturbation series technique is used to compute the quantum state
of light at the output of the fiber. To apply this technique, we
first need to find $G_{\mathrm{hb}}^I(z)$, the interaction-picture
expression of the infinitesimal generator. Using Eqs.~(\ref{GI}) and
(\ref{cont}), one sees that $G_{\mathrm{hb}}^I(z)$ is obtained from
$G_{\mathrm{hb}}(z)$ [Eq.~(\ref{HBG})] by replacing all the
annihilation operators $a_j(z,\pm\Omega)$ by their values
$a_j(0,\pm\Omega)\equiv a_j(\pm\Omega)$ at $z=0$. Note that
$G_{\mathrm{hb}}^I(z)\ne G_{\mathrm{hb}}(0)$.

We now compute the quantum state of light
at the output of the fiber using the first order perturbation
theory: $|\psi(L)\rangle=U^{(1)}(L,0)|0\rangle$, with $U^{(1)}(L,0)$
given by Eq.~(\ref{first}). The result (\ref{FFS}) anticipated in
Sec.~\ref{sec:2} is readily obtained, with
\begin{widetext}
\begin{subequations}\label{HBxi}
\begin{eqnarray}
\xi_{xx}(L,\Omega)&=&i\ \left(\gamma  P_{0x} L\right) \
\e^{2i\theta_{0x}} \
\e^{-i(\beta_2\Omega^2+2\gamma P_{0x})\frac{L}{2}}\ \mathrm{sinc}\left[\left(\beta_2\Omega^2+2\gamma P_{0x}\right)\frac{L}{2}\right],\\
\xi_{yy}(L,\Omega)&=&i\ \left(\gamma  P_{0y} L\right) \
\e^{2i\theta_{0y}} \ \e^{-i(\beta_2\Omega^2+2\gamma
P_{0y})\frac{L}{2}}\
\mathrm{sinc}\left[\left(\beta_2\Omega^2+2\gamma P_{0y}\right)\frac{L}{2}\right],\\
\xi_{xy}(L,\Omega)&=& i \left(\frac{2}{3} \gamma \sqrt{P_{0x}P_{0y}}
L\right)  \e^{i(\theta_{0x}+\theta_{0y})}
\e^{-i[\Delta\beta_1\Omega+\beta_2\Omega^2+\gamma
(P_{0x}+P_{0y})]\frac{L}{2}} \nonumber \\
&\phantom{=}& \times \ \mathrm{sinc}\left[\left(\Delta\beta_1\Omega+\beta_2\Omega^2+\gamma
(P_{0x}+P_{0y})\right)\frac{L}{2}\right], \qquad  \\
\xi_{yx}(L,\Omega)&=&i \left(\frac{2}{3} \gamma \sqrt{P_{0x}P_{0y}}
L\right)  \e^{i(\theta_{0x}+\theta_{0y})}
\e^{i[\Delta\beta_1\Omega-\beta_2\Omega^2-\gamma
(P_{0x}+P_{0y})]\frac{L}{2}} \nonumber \\
&\phantom{=}& \times \ \mathrm{sinc}\left[\left(\Delta\beta_1\Omega-\beta_2\Omega^2-\gamma
(P_{0x}+P_{0y})\right)\frac{L}{2}\right].
\end{eqnarray}
\end{subequations}
\end{widetext}
Eqs.~(\ref{FFS}) and (\ref{HBxi}) describe the FPS
 in high-birefringence optical fibers. Four different
processes can be distinguished according to the polarization of the
generated photons. They are summarized in Tab.~\ref{tab:1}, together
with the corresponding two-photon amplitudes $\xi_{jj'}(L,\Omega )$.
Note that in the processes corresponding to the amplitudes
$\xi_{xx}(L,\Omega )$ and $\xi_{yy}(L,\Omega )$ the pump photons are
co-polarized and have the same polarization as the scattered photons
(scalar scattering). In the processes corresponding to the
amplitudes $\xi_{xy}(L,\Omega )$ and $\xi_{yx}(L,\Omega )$ the pump
photons have orthogonal polarizations (vectorial scattering). Let's
analyze the scalar scattering first.
\begin{table}[h]
  \centering
  \begin{ruledtabular}
  \begin{tabular}{c|c c}
     &  Stokes on $x$ &  Stokes on $y$ \\ \hline
    anti-Stokes on $x$ & $\xi_{xx}(L,\Omega )$ &  $\xi_{xy}(L,\Omega )$\\
    anti-Stokes on $y$ & $\xi_{yx}(L,\Omega )$ & $\xi_{yy}(L,\Omega )$ \\
  \end{tabular}
  \end{ruledtabular}
  \caption{Possible scattering processes in high-birefringence fibers and the corresponding two-photon amplitudes}\label{tab:1}
\end{table}

\subsection{Scalar scattering}\label{sec:4:C}
If the pump wave is linearly polarized along an optical axis of the
fiber (the $x$-axis), only the scalar scattering corresponding to
that axis takes place ($\xi_{yy}(L,\Omega )=\xi_{xy}(L,\Omega
)=\xi_{yx}(L,\Omega )=0$):
\begin{equation}\label{FFSscal}
|\psi^I(L)\rangle=|0\rangle+\int_0^\infty \xi_{xx}(L,\Omega) \
|1^x_\Omega,1^x_{-\Omega}\rangle  \ \d\Omega.
\end{equation}
This formula shows that the generated pairs are in an
\emph{energy-entangled state}: the energies of the signal and idler
photons are correlated while, at the same time, each photon is in a
coherent superposition of a continuum of possible energy
eigenstates.

\begin{figure}[t]
  \centerline{\includegraphics[width=8.5cm]{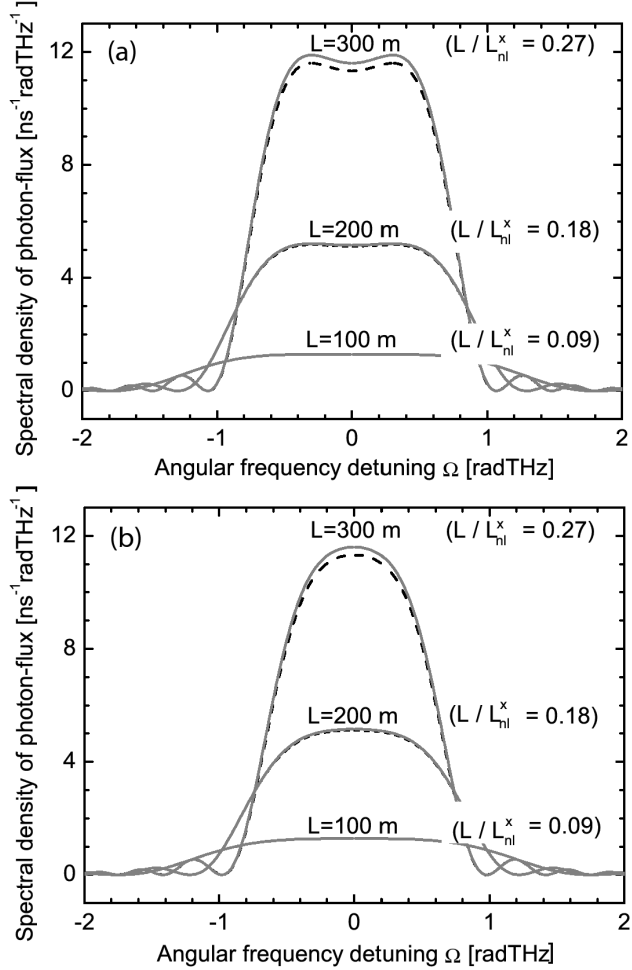}}
  \caption{Photon-flux spectral density $f_x(L,\Omega)$ for the anomalous (a) and normal (b) dispersion regimes. The first-order approximation
  $f_x(L,\Omega)=|\xi_{xx}(L,\Omega)|^2/(2\pi)$ (dashed black lines)
   is compared with the exact value (solid gray lines) obtained by solving Eqs.~(\ref{HBaEvol}) in the Heisenberg picture.
   The parameters used in this figure are $\gamma=3$~/(W~km), $P_{0x}=300$~mW, $\beta_2=\pm 20$~ps$^2$/km
   [$-$ in panel (a) and $+$ in panel (b)], and $L$ takes successively the values 100~m, 200~m, and 300~m.}
   \label{fig:1}
\end{figure}

Let's compute the average photon-flux spectral density
$f_x(L,\Omega)$ at the output of the fiber for an arbitrary angular
frequency $\omega_0+\Omega$. This is given by \cite{brainis2007}
\begin{equation}\label{f}
f_x(L,\Omega)=\frac{1}{2\pi}\lim_{\varepsilon \rightarrow
0}\frac{1}{\varepsilon}
\int_{\Omega-\frac{\varepsilon}{2}}^{\Omega+\frac{\varepsilon}{2}}
\int_{\Omega-\frac{\varepsilon}{2}}^{\Omega+\frac{\varepsilon}{2}} 
 \ \langle
\psi^I(L)|a_x^\dag(\Omega_1)a_x(\Omega_2)|\psi^I(L)\rangle \
\d\Omega_1 \d\Omega_2.
\end{equation}
Using Eqs.~(\ref{FFSscal}) and (\ref{HBxi}a), we find that
\begin{equation}\label{fxapprox}
f_x(L,\Omega)=\frac{|\xi_{xx}(L,\Omega)|^2}{2\pi}=\frac{(\gamma P_{0x}L)^2}{2\pi} \
\mathrm{sinc}^2\left[\left(\beta_2\Omega^2+2\gamma
P_{0x})\right)\frac{L}{2}\right].
\end{equation}
The flux $f_x(L,\Omega)$ is plotted in Fig.~\ref{fig:1} as a
function of the $\Omega$ for a fixed pump power and different fiber
lengths. Panel (a) and (b) are for normal and anomalous dispersion,
respectively. Note that $f_x(L,\Omega)=f_x(L,-\Omega)$ because
signal and idler photons are always created in pairs. The agreement
between the first-order approximation (\ref{fxapprox}) [dashed black
lines] and the exact solution [solid gray line, see Eq.~(\ref{param})] is excellent as long
as the fiber length is much shorter than the non-linearity length:
\begin{equation}\label{Lnl}
L\ll L_{\mathrm{nl}}^x=\frac{1}{\gamma P_{0x}},
\end{equation}
i.e. $f_x(L,\Omega) \ll 1$ for any value of $\Omega$. In this limit,
\emph{the sign of the dispersion influences only slightly the spectral
shape of the created photons} and the spectral width of the
fluorescence spectrum (full first-zero width of the
$\mathrm{sinc}$-function) is $\Delta\Omega_{\mathrm{scal}}\approx 2
\sqrt{2\pi/(|\beta_2|L)}$. The fact that the
sign of $\beta_2$ has almost no influence on the FPS spectrum 
strongly contrasts with the conditions under which scalar modulation 
instability (the stimulated counterpart of scalar FPS) can be observed. The developpement of scalar modulation instability requires  parametric gain, which only
exists in the anomalous dispersion regime (if the dispersion relation is quadratic, as has
been assumed through this work, see Eq.~(\ref{disp})). FPS and the stimulated FWM
phenomena called modulation instabilities will be further compared
in section \ref{sec:6}.

If the pump wave has a finite duration $T$, the
frequency space can be divide in modes of finite spectral width
$\Delta\omega=2\pi/T$ (see the end of Sec.~\ref{sec:2}). Formula
(\ref{fxapprox}) then shows that $|\xi_{xx}(L,\Omega)|^2$ is the mean
number of photons $n_x(L,\Omega)$ in the mode $\omega_0\pm\Omega$. If (\ref{Lnl})
holds, the mean number of photons in any given mode
$\omega+\Omega$ is much lower than one and is therefore numerically
equal to the \emph{probability} of finding a photon in that mode.
The limit $\lim_{T\rightarrow \infty}
n_x(L,\Omega)/T=p_{xx}(L,\Omega)\d\Omega=|\xi_{xx}(L,\Omega)|^2/(2\pi)
\d\Omega$ is the probability per unit of time of finding the photon
in the infinitesimal spectral interval $\d\Omega$. We call
\begin{equation}
p_{xx}(L,\Omega)=\frac{1}{2\pi}|\xi_{xx}(L,\Omega)|^2
\end{equation}
the \emph{spectral density of probability per unit of time} of
creating a photon-pair at the angular frequencies
$\omega_0\pm|\Omega|$. Note that $p_{xx}$ is numerically equal to
$f_x$.

When the condition (\ref{Lnl}) holds, the pump scatters much less
than one photon per mode. However, this condition is not strong
enough to guaranty the validity of Eq.~(\ref{FFSscal}). For
(\ref{FFSscal}) being valid, the \emph{total} probability
$\mathcal{P}_T(L)=T\times \int_0^\infty p_{xx}(L,\Omega) \ \d\Omega$
of scattering a photon during the time $T$ must be much smaller than
one. In order to find an analytical approximation of
$\mathcal{P}_T(L)$, we neglect the term $\gamma P_{0x}L$ in  the
argument of the sinc-function in Eq.~(\ref{fxapprox}) and find that
Eq.~(\ref{FFSscal}) is an accurate approximation of the real quantum state of ligth if
\begin{equation}\label{valid}
\mathcal{P}_T(L)\approx \frac{2}{3} \ (\gamma P_{0x}L)^2
\sqrt{\frac{T^2}{2\pi|\beta_2|L}}\ll 1.
\end{equation}
Since the spectral width of the Stokes (or anti-Stokes) spectrum is
$\Delta\Omega_{\mathrm{scal}}/2 \approx\sqrt{2\pi/(|\beta_2|L)}$ and
the spectral width of the pump is $\Delta\omega=2\pi/T$, the
square-root in Eq.~(\ref{valid}) represents the number of
independent modes in the fluorescence spectrum. Therefore, formula
(\ref{valid}) shows that the expression (\ref{FFSscal}) is a good
approximation of the quantum state of light only when the mean
number of photons per mode is much smaller than the number of
independent modes. However, photon-pair energy-entanglement over a
wide spectral range is rarely desirable. If the Stokes and
anti-Stokes photons are filtered, the condition for having
two-particle energy-entanglement from FPS is that the mean number of
photons per mode must be much smaller than the number of independent
modes in the spectral interval allowed by the filtering process.

\subsection{Vector scattering}\label{sec:4:D}
When the pump wave is not polarized along an optical axis, not only
scalar but also vectorial scattering can take place. The quantum
state of light at the output of the fibre is then given by
Eq.~(\ref{FFS}). The two-photon amplitudes $\xi_{ij}(L,\Omega)$,
with $(i,j)\in\{x,y\}^2$, are given in Eqs.~(\ref{HBxi}). Amplitudes
$\xi_{xx}$ and $\xi_{yy}$ correspond to a scalar scattering on the
$x$ and $y$ axes, respectively. The remaining two amplitudes
$\xi_{xy}$ and $\xi_{yx}$ correspond to a vectorial scattering in
which two pump photons with orthogonal polarizations are annihilated.
If we chose $\Delta\beta_1>0$ (convention), the $x$-axis is the slow
axis. Therefore (see Tab.~\ref{tab:1}), $\xi_{xy}$ corresponds to
the scattering of the most energetic photon of the pair (the
anti-Stokes photon) on the slow axis, while $\xi_{yx}$ corresponds
to the opposite process where the most energetic photon is polarized
along the fast axis.

As in the scalar case, the final state $|\psi^I(L)\rangle$ is an
energy-entangled state. However, some polarization-entanglement is
also present. The kind of polarization-entanglement that can be
generated using HB fibers depends on the spectral shapes of the
two-photon amplitudes, that in turn depend on the fiber parameters,
the power of the pump, and its polarization state. We investigate
the possible cases by first calculating the average spectral densities of
photon flux, $f_x(L,\Omega)$ and $f_y(L,\Omega)$, generated on the
$x$ and $y$ axes, respectively.
\subsubsection{Spectral density of photon-flux}
 Using Eq.~(\ref{f}) (and a similar
relation for the $y$-axis), one finds
\begin{subequations}\label{fluxVMI}
\begin{eqnarray}
f_x(L,\Omega)&=&\frac{1}{2\pi}\left(|\xi_{xx}(L,\Omega)|^2+\Theta(\Omega) \ |\xi_{xy}(L,\Omega)|^2\right. \nonumber \\
&\phantom{=}&+\left. \Theta(-\Omega) \ |\xi_{yx}(L,-\Omega)|^2\right), \\
f_y(L,\Omega)&=&\frac{1}{2\pi}\left(|\xi_{yy}(L,\Omega)|^2+\Theta(\Omega) \ |\xi_{yx}(L,\Omega)|^2\right. \nonumber \\
&\phantom{=}&+\left. \Theta(-\Omega) \
|\xi_{xy}(L,-\Omega)|^2\right),
\end{eqnarray}
\end{subequations}
where $\Theta(\Omega)=0$ for $\Omega<0$ and $\Theta(\Omega)=1$ for
$\Omega>0$.

\begin{figure}[t]
  \centerline{\includegraphics[width=8.5cm]{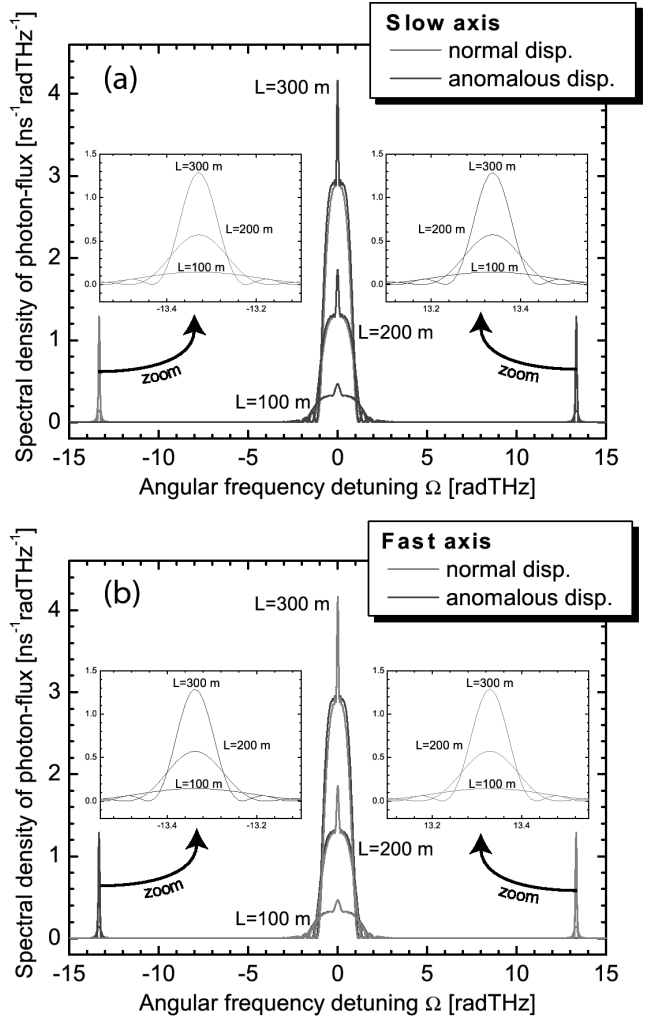}}
  \caption{Photon-flux spectral densities $f_x(L,\Omega)$ (a) and $f_y(L,\Omega)$ (b) as a function of $\Omega$ for
  propagation distances $L=100$~m, $L=200$~m, and $L=300$~m. The other parameters used in this figure are $\gamma=3$~/(W~km), $P_{0x}=P_{0y}=150$~mW, $\Delta\beta_1=200$~ps/km, $\beta_2=\pm 15$~ps$^2$/km. Light gray and dark gray curves correspond to the
   normal ($\beta_2=+ 15$~ps$^2$/km) and anomalous dispersion ($\beta_2=- 15$~ps$^2$/km), respectively.}
   \label{fig:2}
\end{figure}

The functions $f_x(L,\Omega)$ and $f_y(L,\Omega)$ are plotted in
Figs.~\ref{fig:2}(a) and \ref{fig:2}(b), respectively, for typical
parameters of HB silica fibers and an equipartition of the total
pump power $P_0$ between the optical axes ($P_{0x}=P_{0y}=P_0/2$).
Light gray and dark gray curves correspond to normal and anomalous
dispersion; the other parameters are otherwise the same, including
the absolute value of $\beta_2$. Three propagation lengths are
considered. In all cases, $L\ll
L_{\mathrm{nl}}=\min[L_{\mathrm{nl}}^x,L_{\mathrm{nl}}^y]$, where
$L_{\mathrm{nl}}^j=1/(\gamma P_{0j})$ in the non-linearity length
corresponding to the $j$-axis, so that the first-order perturbation
formulas (\ref{fluxVMI}) are accurate approximations.
An important feature of Fig.~\ref{fig:2} is that the scalar and
vector FPS take place in well separated frequency ranges. The broad
central peak around $\Omega=0$ is due to scalar scattering, as
described in Sec.~\ref{sec:4:C}. Two separate scalar FPS
processes take place of the slow and fast axes. The narrow spectral
peaks at higher frequency detuning
\begin{equation}\label{HBpeakpos}
|\Omega_{\mathrm{vect}}|\approx\frac{\Delta\beta_1}{|\beta_2|}\left(1-\alpha\right)
\end{equation}
are due to the vector scattering. In Eq.~(\ref{HBpeakpos}),
\begin{equation}\label{alpha}
\alpha =\frac{\beta_2\times \gamma P_0}{\Delta\beta_1^2}.
\end{equation}
It has been shown in \cite{nguyen2006} that the dynamics of the
fields in a HB fiber can be discussed in terms of this single
dimensionless parameter. The situation displayed in Fig.~\ref{fig:2}
correspond to $\alpha\ll 1$. As seen from the figure, 
\emph{changing the sign of the
group-velocity dispersion has the same effect as
interchanging $\Omega$ and $-\Omega$}. The narrow central peak at
$\Omega=0$ is also due to vector scattering, but being positioned at
the pump frequency, it is difficult to observe in practice. Only one of the amplitudes $\xi_{xy}(L,\Omega)$
and $\xi_{yx}(L,\Omega)$ takes significant values for $\Omega>0$
[see the definitions (\ref{HBxi}c) and (\ref{HBxi}d)]. In the normal
dispersion regime, vector FPS is due to non-zero values of
$\xi_{yx}$ in the neighborhood of the angular frequency
(\ref{HBpeakpos}). In the anomalous dispersion regime, $\xi_{yx}$ is zero for any $\Omega>0$; $\xi_{xy}$ takes significant values
in the neighborhood of the angular frequency (\ref{HBpeakpos}),
instead. Tab.~\ref{tab:1} explains the polarization properties of
the scattered photons. The spectral width of the vector FPS peaks is
$\Delta\Omega_{\mathrm{vect}}\approx4\pi/(\Delta\beta_1 L)$. As
illustrated in Fig.~\ref{fig:2}, \emph{very monochromatic} ($\Delta\Omega_{\mathrm{vect}}\approx0.2$~radTHz) photon pairs can be generated using vectorial FPS. With picosecond pump pulses,  \emph{Fourier-transform limited photon pairs could be generated without any external filtering scheme}.  

\begin{figure}[t]
  \centerline{\includegraphics[width=8.5cm]{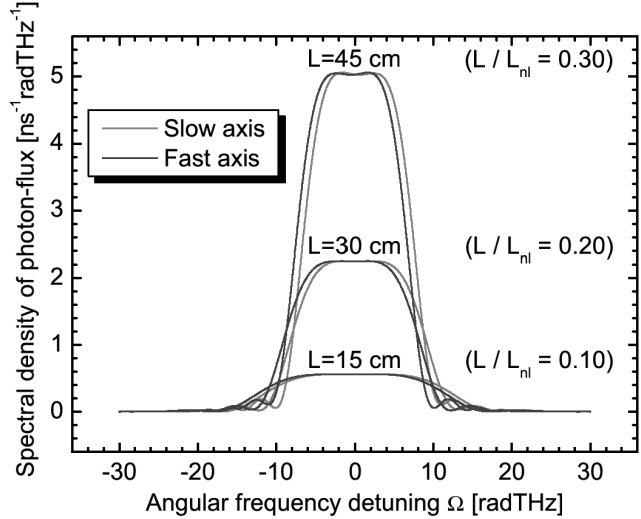}}
  \caption{Photon-flux spectral densities $f_x(L,\Omega)$ (light gray) and $f_y(L,\Omega)$ (dark gray) as a function of $\Omega$
  for propagation distances $L=15$~cm, $L=30$~cm, and $L=45$~cm. The other
parameters used in this figure are $\gamma=36$~/(W~km),
$P_{0x}=P_{0y}=20$~W, $\Delta\beta_1=400$~ps/km, $\beta_2=-
139$~ps$^2$/km ($\alpha=-1.25$).}
   \label{fig:3}
\end{figure}

As the value of the parameter $|\alpha|$ increases, the frequency ranges of
the scalar and vector scattering become closer and can even merge. In the $P_{0x}=P_{0y}=P_{0}/2$ case, this happens
for $|\alpha|\ge 1$. This behavior has been recently observed in a
photonic-crystal fiber \cite{nguyen2006} in a modulation instability
regime. Fig.~\ref{fig:3} shows the spectral
densities of photon-flux for a fiber with parameters similar to
those in experiment \cite{nguyen2006}, but in a regime where $L\ll
L_{\mathrm{nl}}=\min[L_{\mathrm{nl}}^x,L_{\mathrm{nl}}^y]$ holds.
Scalar and vector scattering occur in the same frequency range. One
cannot distinguish photons from these different processes by their
energy anymore. 

\subsubsection{Polarization-entanglement}
Energy-entanglement is a feature common to any, $\chi^{(2)}$ or
$\chi^{(3)}$, parametric photon-pair generation process. In
contrast, intrinsic polarization-entanglement is only encountered in
special circumstances. In $\chi^{(2)}$ parametric down-conversion,
for instance, it requires a type-II phase-matching \cite{kwiat1995}.
In this paragraph, the kind of polarization-entanglement achievable
using the FPS process in HB fibers is described. To separate
polarization-entanglement from energy-entanglement, let us consider
that the light coming out of the fiber is processed through a
filtering apparatus that only selects, from the full spectrum, one
pair of correlated modes with frequencies $\omega_0\pm\Omega$ and
spectral width $\Delta\omega=2\pi/T$ (where $T$ is, as before, the
pump duration). We also consider, for simplicity, that
$P_{0x}=P_{0y}=P_{0}/2$.

Let's first consider the $|\alpha|\ll 1$ case illustrated in
Fig.~\ref{fig:2}. If $\Omega$ corresponds to the center of the
narrow vector scattering peaks, the quantum state of the generated
photon pair is
\begin{equation}\label{state1}
|1_\Omega^y 1_{-\Omega}^x\rangle
\end{equation}
for normal dispersion, and
\begin{equation}\label{state2}
|1_\Omega^x 1_{-\Omega}^y\rangle
\end{equation}
for anomalous dispersion. The generated photons always have opposite
polarizations but they \emph{are not} entangled. The polarization of 
the photons is correlated to their energy.

Consider now that $\Omega$ is somewhere in the frequency range
corresponding to the scalar scattering. In that case, the generated
photon pair will be in the quantum state
\begin{equation}\label{bell}
\frac{1}{\sqrt{2}}\left(|1_\Omega^x 1_{-\Omega}^x\rangle +
\e^{i\phi} \ |1_\Omega^y 1_{-\Omega}^y\rangle\right),
\end{equation}
where
\begin{equation}
\phi=2(\theta_{0y}-\theta_{0x})
\end{equation}
The state (\ref{bell}) is a maximally entangled two-particle state
(Bell state). The relative phase $\phi$ only depends on the
polarization of the pump field. It can be controlled using a
quarter-wave plate placed in the pump beam. Since the polarization
of each photon of the pair can also be rotated individually after they exited the fiber, any Bell
state can be produced. If the pump powers propagating on the optical
axes are not equal ($P_{0x}\neq P_{0y}$), the two components of the
coherent superposition (\ref{bell}) do not have the same amplitude.
In that case, the state is not maximally entangled. The fraction of
the total pump power $P_0$ that is polarized along the slow axis can
be controlled using a half-wave plate. Therefore, a full control of
the pump polarization (quarter-wave and half-wave plates) permits to
modify the relative phase and the strength of the entanglement.

Consider now that the spectral range of scalar and vector scattering
overlap. If the photons in the modes $\omega_0\pm\Omega$ can be
produced by either the scalar or the vector process, the amount of
entanglement will be reduced. In the extreme case depicted in
Fig.~\ref{fig:3} ($|\alpha|\ge 1$), the state
\begin{equation}
\left(|1_\Omega^x 1_{-\Omega}^x\rangle + \e^{i\phi} \ |1_\Omega^y
1_{-\Omega}^y\rangle\right) + 2 \times \left(|1_\Omega^x \rangle+ \e^{i\frac{\phi}{2}} \
|1_\Omega^y\rangle\right)\left(|1_{-\Omega}^y\rangle +
\e^{i\frac{\phi}{2}} \ |1_{-\Omega}^x\rangle\right)
\end{equation}
is produced if $\Omega$ is small enough (plateau in Fig.~\ref{fig:3}). This state is partially entangled.

Generating any possible polarization-entangled Bell states
(\ref{bell}) in a HB optical fiber is a very attractive prospect for
quantum information processing. A first experimental demonstration
has been realized very recently \cite{Fan2007c,Fan2007b} (in a
counter-propagating geometry to avoid any walk-off of the $x$- and
$y$-polarized photons). The other processes described in this
section have not been demonstrated yet. The process that producing
the states (\ref{state1}) and (\ref{state2}) is very interesting
from an experimental point of view because it generates photon pairs
that are very narrow-band (see Fig.~\ref{fig:2}) and easy to separated (at the fiber end) using the polarization
degree of freedom.

\section{Low-birefringence fibers}\label{sec:5}
We now examine the peculiarities of FPS in fibers having a low
birefringence (LB fibers). In that case, the group-velocity mismatch
can be neglected ($\beta_{1x}=\beta_{1y}\equiv\beta_{1}$). In
contrast with the HB case, the last terms in the right-hand side of
Eqs.~(\ref{QNLS}) cannot be dropped because the beat length
$2\pi/\Delta\beta_0$ may have the same order of magnitude as the
other relevant length scale (the non-linearity length, for
instance). The propagation equations are
\begin{subequations}\label{LBQNLS}
\begin{eqnarray}
\frac{\partial A_x}{\partial z}+\beta_{1}  \frac{\partial
A_x}{\partial t}+i  \frac{\beta_2}{2}  \frac{\partial^2
A_x}{\partial t^2}&=& i \gamma  \left(A_x^\dag
A_x^{\phantom{\dag}}+\frac{2}{3}
A_y^\dag A_y^{\phantom{\dag}}\right)A_x\nonumber\\
&+&
i  \frac{\gamma}{3}  A_y^2A_x^\dag   \e^{-2i\Delta\beta_0z},\\
\frac{\partial A_y}{\partial z}+\beta_{1}  \frac{\partial
A_y}{\partial t}+i  \frac{\beta_2}{2}  \frac{\partial^2
A_y}{\partial t^2}&=& i  \gamma  \left(A_y^\dag
A_y^{\phantom{\dag}}+\frac{2}{3}
A_x^\dag A_x^{\phantom{\dag}}\right)A_y\nonumber \\
&+&i  \frac{\gamma}{3}  A_x^2A_y^\dag \e^{+2i\Delta\beta_0z}.
\end{eqnarray}
\end{subequations}

As explained at the end of Sec.~\ref{sec:3}, only FPS due to a
monochromatic pump wave \emph{polarized along an optical axis} will
be studied here. We arbitrary choose this axis as the $x$-axis. The
cartesian components of the envelope of the injected monochromatic
pump are
\begin{equation}\label{LBini}
A_x(0,t)=\sqrt{P_0} \ \e^{i\theta_{0}}; \qquad A_y(0,t)=0.
\end{equation}
Whether the $x$-axis is the slow or fast axis depends on the sign of
$\Delta \beta_0$. The monochromatic solution of Eqs.~(\ref{LBQNLS})
that is compatible with the initial condition (\ref{LBini}) is
\begin{equation}\label{LBclas}
A_x(z,t)=\sqrt{P_0} \ \e^{i\theta_{0}}\ \e^{i\gamma P_0 z}; \qquad
A_y(z,t)=0.
\end{equation}

As in the high-birefringence case, we want to linearize the
propagation equations in the neighborhood of the monochromatic
solution (\ref{LBclas}). Therefore, we introduce the ansatz
\begin{subequations}\label{LBansatz}
\begin{eqnarray}
A_{x}(z,t)&=& \left(\sqrt{P_0}+u_{x}(z,t)\right) \e^{i\theta_0} \
\e^{i \gamma P_0 z}, \\
A_{y}(z,t)&=& u_{y}(z,t) \ \e^{i \frac{2}{3}\gamma P_0 z}
\end{eqnarray}
\end{subequations}
and inject it in (\ref{LBQNLS}). Linearizing the propagation
equations with respect to $u_x$ and $u_y$ results in:
\begin{subequations}\label{LBpert}
\begin{eqnarray}
\frac{\partial u_x}{\partial z}&=&-\beta_{1}\frac{\partial
u_x}{\partial t}-i \frac{\beta_2}{2} \frac{\partial^2 u_x}{\partial
t^2}+i\ \gamma \
P_{0}\left(u_x+u_x^\dag\right)\qquad \\
\frac{\partial u_y}{\partial z}&=&-\beta_{1}\frac{\partial
u_y}{\partial t}-i \frac{\beta_2}{2} \frac{\partial^2 u_y}{\partial
t^2}\nonumber \\
&\phantom{=}&+i \ \frac{\gamma}{3} \ P_{0} \ u_y^\dag \
\e^{2i\theta_0} \ \e^{2i(\Delta\beta_0+\frac{1}{3}\gamma P_0)z}.
\end{eqnarray}
\end{subequations}
These equations show that the perturbations $u_x$ and $u_y$ are
\emph{uncoupled}. Consequently, no correlation is expected between
photons generated on the $x$- and $y$-axes.
\subsection{Coupled-mode equations}
Using Eqs.~(\ref{cpm}), (\ref{modal1}), (\ref{A}) and
(\ref{LBansatz}), the mode expansion of the fields $u_x$ and $u_y$
is obtained:
\begin{subequations}\label{LBdecomp}
\begin{eqnarray}
u_x(z,t)&=&\sqrt{\frac{\hbar\omega_0}{ 2\pi}}
 \  \e^{-i\theta_{0}} \  \e^{i\gamma
 P_{0}z}\nonumber \\
&\phantom{=}& \times \int  \ a_j(z,\Omega)
\ \e^{iB(\Omega)z-i\Omega t} \ \d\Omega,\\
u_y(z,t)&=&\sqrt{\frac{\hbar\omega_0}{ 2\pi}} \ \int a_j(z,\Omega) \
\e^{iB(\Omega)z-i\Omega t} \d\Omega,
\end{eqnarray}
\end{subequations}
with $B(\Omega)=\beta_1\Omega+(\beta_2/2)\Omega^2$. Injecting the
expansions (\ref{LBdecomp}) in Eqs.~(\ref{LBpert}), we obtaine the following
coupled-mode equations:
\begin{subequations}\label{LBaEvol}
\begin{eqnarray}
\frac{\partial}{\partial z} a_x(z,\Omega)&=& i \ \gamma  \ P_0 \
\e^{i2\theta_0}  \nonumber \\
&\phantom{=}& \times \  \e^{-i(\beta_2\Omega^2+2\gamma P_0)z}\
a^\dag_x(z,-\Omega)\\
\frac{\partial}{\partial z} a_y(z,\Omega)&=& i \  \frac{\gamma}{3} \
P_0 \ \e^{i2\theta_0} \nonumber \\
&\phantom{=}& \times \ \e^{-i(\beta_2\Omega^2-\frac{2}{3} \gamma
P_0-2\Delta\beta_0)z} \ a^\dag_y(z,-\Omega). \ \qquad
\end{eqnarray}
\end{subequations}
Note that the first equation is the same as equation
(\ref{HBaEvol}a) for $P_{0y}=0$. We know, from the study of
Sec.~\ref{sec:4:C}, that this corresponds to a scalar scattering on
the $x$-axis, and will not repeat this analysis here. The
interesting feature of FPS in LB fibers comes from
Eq.~(\ref{LBaEvol}b), which described a new kind of vectorial FPS,
as can be seen from the structure of the infinitesimal generator:
\begin{widetext}
\begin{eqnarray}\label{LBG}
G_{\mathrm{lb}}(z)&=& \frac{\hbar}{2}\gamma P_0
\left(\e^{2i\theta_{0}}\int \e^{-i(\beta_2\Omega^2+2\gamma P_0) z} \
a_x^\dag(z,\Omega) a_x^\dag(z,-\Omega)
\ \d\Omega + \mathrm{H.c.}\right) \nonumber\\
&\phantom{=}&+ \frac{\hbar}{2} \frac{1}{3}\gamma P_0
\left(\e^{2i\theta_{0}}\int \e^{-i(\beta_2\Omega^2-\frac{2}{3}\gamma
P_0-2\Delta\beta_0) z} \ a^\dag_y(z,\Omega) a_y^\dag(z,-\Omega) \
\d\Omega + \mathrm{H.c.}\right).
\end{eqnarray}
\end{widetext}
\subsection{Four-photon scattering}
Using the infinitesimal generator (\ref{LBG}) and moving to the
interaction picture, the quantum state of light at the output of the
fiber is found to be
\begin{eqnarray}\label{FFSlb}
|\psi^I(L)\rangle&=&|0\rangle+\int_0^\infty \xi_{xx}(L,\Omega) \
|1^x_\Omega,1^x_{-\Omega}\rangle  \ \d\Omega \nonumber \\
&\phantom{=}&+\int_0^\infty \xi_{yy}(L,\Omega) \
|1^y_\Omega,1^y_{-\Omega}\rangle  \ \d\Omega,
\end{eqnarray}
where $\xi_{xx}$ is given by Eq.~(\ref{HBxi}a) (with $P_{0x}=P_0$)
and
\begin{eqnarray}\label{LBxi}
\xi_{yy}(L,\Omega)&=&i\ \left(\gamma  \frac{P_{0}}{3} L\right) \
\e^{2i\theta_{0}} \ \e^{-i(\beta_2\Omega^2-\frac{2}{3}\gamma
P_0-2\Delta\beta_0)\frac{L}{2}} \nonumber \\
&\phantom{=} &\times
\mathrm{sinc}\left[\left(\beta_2\Omega^2-\frac{2}{3}\gamma
P_0-2\Delta\beta_0\right)\frac{L}{2}\right].
\end{eqnarray}
The $x$-polarized pump not only scatters $x$-polarized photons
through the scalar scattering process but also $y$-polarized photons
through a vectorial scattering process. This vectorial process is
different from those encountered in the study of FPS in HB fibers:
here two linearly co-polarized pump photons give birth to a
signal-idler pair polarized \emph{orthogonally} to them.

\begin{figure}[t]
  \centerline{\includegraphics[width=8.5cm]{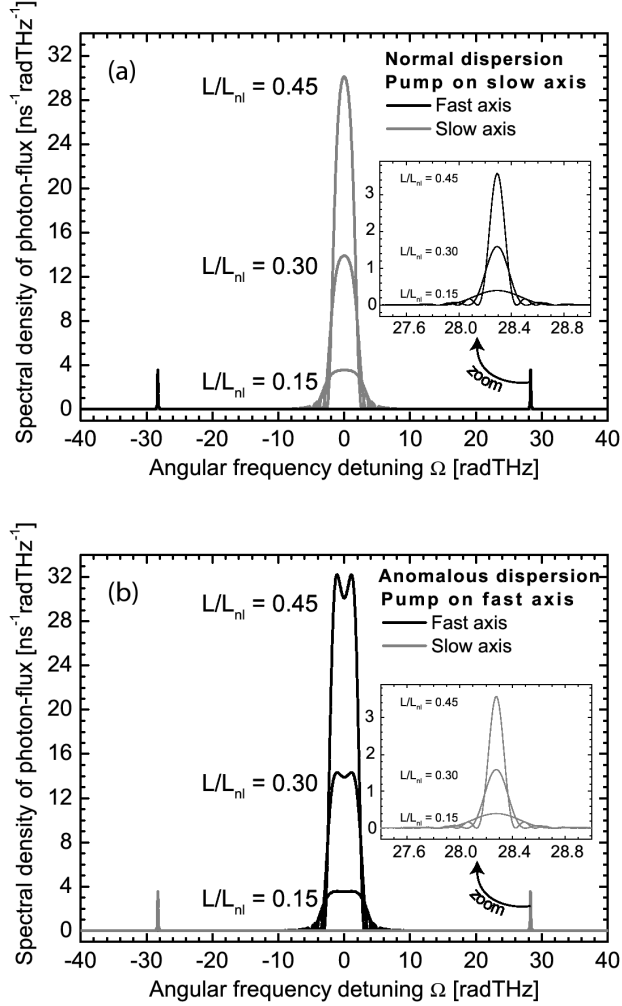}}
  \caption{Photon-flux spectral densities on the slow (gray lines) and fast (black lines)  axes of the fiber
  as a function of $\Omega$. In the panel (a), the dispersion is
  normal and the pump in polarized along the slow axis of the fiber.
  In the panel (b), the situation is opposite: the dispersion in
  anomalous and the polarization of the pump wave is along the fast
  axis. The
parameters used in this figure are $\gamma=3$~/(W~km), $P_{0}=1$~W,
$|\Delta\beta_0|=2$~m$^{-1}$, and $|\beta_2|=5$~ps$^2$/km. Three
propagation distances ($L=50$~m, $L=100$~m, and $L=150$~m) are
considered. }
   \label{fig:4}
\end{figure}

The spectral density of photon flux associated with scattering in a
low-birefringent fiber in shown in Fig.~\ref{fig:4}.
In Fig.~\ref{fig:4}a, $\beta_2>0$ and the pump is linearly polarized
along the slow axis ($\Delta\beta_0>0$). The central peak is due to
scalar scattering. The two narrow far-detuned peaks polarized along
the fast axis are due to the vector FPS associated to the amplitude
(\ref{LBxi}). Note that if the pump is polarized along the fast axis
($\Delta\beta_0<0$), an anomalous dispersion regime ($\beta_2<0$) is
needed to observe a similar scattering on the slow axis, as
displayed in Fig.~\ref{fig:4}b. In both cases, the detuning and the spectral width (full width at the first-zeros of the sinc-function) of the generated photon pairs are $|\Omega_{\mathrm{vect,lb}}|\approx\sqrt{2\Delta\beta_{0}/\beta_{2}}$ and $\Delta\Omega_{\mathrm{vect,lb}}\approx(2\pi/L) /\sqrt{2 \beta_{2}\Delta\beta_{0}}$, respectively. When $\Delta\beta_0\times
\beta_2<0$, FPS only occurs at frequency detunings close to
$\Omega=0$. This case is not shown in Fig.~\ref{fig:4}.

Photon-pair generation using vector FPS in LB fibers has not been
demonstrated yet. Vector FPS in LB fibers has however an important
advantage over scalar FPS in terms of signal-to-noise ratio.  It
has been shown that, photon-pair correlations can be one order of
magnitude higher than in the scalar FPS case for angular frequency
detunings $|\Omega|$ between 30 and 90 radTHz \cite{Lin2006}.
Furthermore, photons generated by vector FPS are very narrow-band: as shown in Fig.~\ref{fig:4}, $\Delta\Omega_{\mathrm{vect,lb}}\approx$~0.2~THz. By pumping with picosecond pulses, Fourier-transform limited photon pairs could be produced without filtering. Monochromaticity and low Raman noise are the two advantages offered by vectorial FPS in LB fibers.

\section{Connection with modulation instabilities}\label{sec:6}
There is a connection between FPS from a single pump beam and the
phenomenon called ``modulation instability'' in nonlinear fiber
optics \cite{agraw}. In this work, a special care has been taken to
formulate the perturbation theory of FPS in a way that makes this
connection obvious.

A modulation instability is a nonlinear phenomenon that allows an
initially continuous monochromatic wave to become modulated in time
due to the propagation in the fiber. This effect can be seen as the
result of a degenerate four-wave mixing (FWM) process that is self
phase-matched by a balanced between birefringence, dispersion and
non-linearity. The pump develops exponentially growing symmetric
sidebands about the frequencies $\omega_0\pm\Omega_m$ satisfying
phase-matching conditions. In the time domain, the envelope of the
pump is therefore modulated at angular frequency $\Omega_m$.

The dynamics of modulation instabilities 
 can be understood by solving the equation systems
(\ref{HBaEvol}) or (\ref{LBaEvol}) exactly. This method is used in
classical nonlinear optics where $a_j(z,\Omega)$ and
$a_j^\dag(z,\Omega)$ are treated as classical spectral amplitudes.
The analysis provides the phase-matching conditions, the frequency
range of the instable (i.e. exponentially growing) modes
experiencing the parametric gain, the value of that gain, and the
polarization of the growing sidebands. The development of the
instability can be induced by a additional coherent probe signal at
$\omega_0\pm\Omega_m$ or even some incoherent optical noise
travelling with the pump. If care is taken to eliminate these
instability sources, modulation instability can still develop
because the photon pairs produced by the FPS phenomenon populate the
unstable modes. In that case, the modulation instability is called
\emph{spontaneous}, and has a purely quantum origin.

It is important to note that although modulation instabilities can
be trigger by the FPS process, their spectral properties are
different. Let us illustrate that point in the case of a scalar
modulation instability. The analysis of Eqs.~(\ref{HBaEvol}) with
$P_{0x}=P_0$ and $P_{0y}=0$ shows that instability is only observed
in anomalous dispersion regime ($\beta_2<0$). (In reality, an
instability can also be observed in the normal dispersion regime if
the pump wavelength is close to the zero-dispersion wavelength; this
results from higher order dispersion effects that have been omitted
here by limiting the Taylor expansion (\ref{disp}) to second order.)
The parametric gain $g(\Omega)=\sqrt{(\gamma P_0)^2-(2\gamma
P_0-|\beta_2|\Omega^2)^2/4}$ only exists in the angular frequency
range $]-2\sqrt{\gamma P_0/|\beta_2|},2\sqrt{\gamma
P_0/|\beta_2|}[$. Its maximum value $g_{\mathrm{max}}=\gamma P_0$ is
reached for the angular frequencies
$\Omega_{\mathrm{max}}=\pm\sqrt{2\gamma P_0/|\beta_2|}$
corresponding to a perfect phase-matching with the pump. For
$g(\Omega)L\gg1$, the photon flux that results from a spontaneous
scalar modulation instability is well approximated by
\begin{equation}\label{gainMI}
f_x(L,\Omega)=\frac{1}{2\pi} \frac{\gamma^2 P_0^2}{4 g^2(\Omega)} \
\e^{2g(\Omega) L}.
\end{equation}
Because of the exponential growth, the spectrum associated with
modulation instability exhibits sharp peaks at detunings
$\pm\Omega_{\mathrm{max}}$. This strongly contrasts with the broad
fluorescence spectrum due to scalar FPS. The build-up of the
modulation instability can be understood by comparing the spectral
intervals over which parametric amplification and FPS take place.
For a given pump power $P_0$, the width of the spectral interval in
which scalar modulation instability develops is
$\Delta\Omega_{\mathrm{SMI}}=4\sqrt{\gamma P_0/|\beta_2|}$. It is
independent of the propagation length. In contrast, FPS takes place
in a spectral interval $\Delta\Omega_{\mathrm{scal}}\approx 2
\sqrt{2\pi/(|\beta_2|L)}$, that depends on the propagation length
$L$ but is independent of the pump power to a good approximation
(see Sec.~\ref{sec:4:C}). The ratio of these bandwidth is
$r=\Delta\Omega_{\mathrm{SMI}}/\Delta\Omega_{\mathrm{scal}}\approx
0.8 \times \sqrt{\gamma P_0 L}$. This shows that, in the beginning
of the propagation ($\gamma P_0 L\ll 1$), the FPS spectrum is much
broader that the amplification band. As light propagates further,
the number of photons per mode increases and the FPS spectrum
becomes narrower. When $\gamma P_0 L \approx 1$ (the propagation
length is equal to the non-linearity length
$L_{\mathrm{nl}}=1/\gamma P_0$), the width of the FPS spectrum is
almost equal to the width of the parametric amplification band. At
that moment, each mode in the amplification band has been populated
with about one photon per mode. Stimulated FWM can therefore take
place and modulation instability develops.

\begin{figure}[t]
  \centerline{\includegraphics[width=8.5cm]{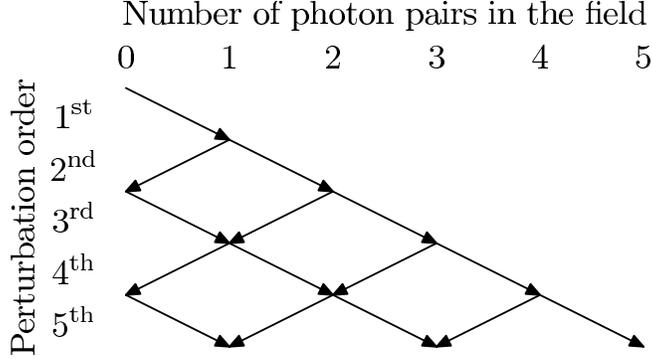}}
  \caption{Graphical representation of the mechanism leading to the growth of the number of photon pairs in the field
  in the
  case of a scalar scattering. Each arrow represents a scattering event, i.e.
  the creation (right-pointing arrow)
  or the annihilation (left-pointing arrow) of a signal/idler pair.
  Column's labels represent the number of photon pairs in the field.
  Rows' labels indicate the order in the Dyson's perturbation
  series accounting for the scattering events represented by arrows in that row.}
   \label{fig:5}
\end{figure}

The discussion above is qualitative because, when the mean number of
photons per mode approaches one, the scattering theory based on the
first-order perturbation approach (FPS) is not accurate anymore. To
describe what happens when the number of photons scattered from the
pump is higher than one, higher-order perturbation theories based on
Eq.~(\ref{norder}) with $n>1$ can be used. Fig.~\ref{fig:5}
shows how the different perturbation orders (\ref{norder}) in
Dyson's series contribute to the quantum state of light in the case
of scalar scattering.

 The rows represent successive perturbation orders, while the
columns represent the number of photon pairs in the field. An arrow
connecting a node with $n$ photon pairs at perturbation order $k$ to
a node with $l$ photon pairs at perturbation order $k+1$ means that
the amplitude of the $l$-pair state (at order $k+1$) depends on the
amplitude of the $n$-pair state (at order $k$). The figure shows
that limiting the Dyson's series to the $k^{\mathrm{th}}$ order
consists in approximating the quantum state of light by a state
containing no more that $k$ photon pairs. This approximation is
therefore precise if the probability of having $k$ photon pairs in
the field is small. It also shows that the amplitude of a $n$-pair
state at a given perturbation order $k$ only depends on the
amplitudes of the states with $n-1$ and $n+1$ photon pairs at order
$k-1$. For instance, in the second order approximation, the quantum
state is
\begin{eqnarray}\label{secorder}
|\psi\rangle&=&[1-a(L)] |0\rangle+\int_0^\infty \xi_{xx}(L,\Omega) \
|1^x_\Omega,1^x_{-\Omega}\rangle  \ \d\Omega \nonumber \\
&\phantom{=}&+ \frac{1}{2}\left(\int_0^\infty \xi_{xx}(L,\Omega) \
|1^x_\Omega,1^x_{-\Omega}\rangle  \ \d\Omega\right)^2.
\end{eqnarray}
The term in the second line of (\ref{secorder}) represents the
possibility of emitting two photon pairs, and $a(L)$ is the
second-order correction to the amplitude of the initial ``no
photon-pair'' state $|0\rangle$. The probability of creating at
least one photon pair is equal to $2\Re[a(L)]$. This number only
takes a finite value if the pump wave has a finite duration $T$. For
$\gamma P_{0x} L\ll 1$, $2\Re[a(L)]\approx \mathcal{P}_T(L)$ [see
Eq.~(\ref{valid})]. One can easily compute that the mean number of
photons generated during the time $T$ in a spectral mode around the
frequency detuning $\Omega$ is
\begin{equation}
n_x(L,\Omega)=|\xi_{xx}(L,\Omega)|^2+|\xi_{xx}(L,\Omega)|^2\mathcal{P}_T(L)+|\xi_{xx}(L,\Omega)|^4.
\end{equation}
The first term corresponds to the first-order approximation while
the second and third terms are second-order corrections. The second
term describes a second independent FPS event. The probability of
scattering a second photon in the mode $\Omega$ is equal to the
probability of scattering a first photon (regardless the mode) times
the probability of scattering a photon in the mode $\Omega$. The
third term is interpreted as a stimulated emission contribution. An additional
photon in the $\Omega$ mode is more likely if the first one as
already been emitted in that mode.

Higher-order perturbation theory provides important insights to the
physics of photon-pair generation in fibers and the built-up of
spontaneous modulation instabilities. However, it is clearly not an
efficient way to compute the mean number of photons in a given
spectral mode. In the case of scalar scattering, the quantum
equations of motion (\ref{HBaEvol}), with $P_{0x}=P_0$ and
$P_{0y}=0$, can be solved analytically without any approximation.
This solution is equivalent to summing the Dyson's series. The
average photon-flux spectral density derived by that method is
\begin{equation}\label{param}
f_x(L,\Omega)=\frac{1}{2\pi} \frac{\gamma^2
P_0^2}{|\lambda(\Omega)|^2} \
\left|\sinh(\lambda(\Omega)L)\right|^2,
\end{equation}
where $\lambda(\Omega)=\sqrt{(\gamma P_0)^2-(2\gamma
P_0+\beta_2\Omega^2)^2/4}$. In the limit $\gamma P_0 L \ll 1$, the
formula (\ref{param}) reduces to (\ref{fxapprox}). In the anomalous
dispersion regime ($\beta_2<0$) and the $\gamma P_0 L \gg 1$ limit,
it gives Eq.~(\ref{gainMI}); note that $\lambda(\Omega)$ can only be
identified with the gain $g(\Omega)$ when
$\lambda(\Omega)\in\mathbb{R}$.

Eq.~(\ref{param}) is a well known result for a parametric
$\chi^{(3)}$ amplifier \cite{Wang2001}. In the case of vectorial
scattering, the quantum equations of motion, Eqs.~(\ref{HBaEvol}) in
the HB case or Eqs.~(\ref{LBaEvol}) in the LB case, can be solved
without approximations only when only one scattering process
contributes to the population of the signal ($-\Omega$) and idler
($\Omega$) modes under consideration.
For these cases, formulas similar to (\ref{param}) can then be
derived. In HB fibers, different processes can scatter pump photons
to the same spectral interval. As seen in Sec.~\ref{sec:4:D}, this
happens when the parameter $|\alpha|$ [Eq.~(\ref{alpha})] is higher
than one. In that case, Eqs.~(\ref{HBaEvol}) cannot be solved
analytically. Perturbation theory, however, permits to study these
cases. From a general point of view, solving the equations of motion
exactly is not a good theoretical approach to the photon-pair
generation problem because the physical process of FPS cannot be
straightforwardly derived from the mathematical solution.
First-order perturbation theory, on the other hand, clearly exhibits
the physical process of FPS, applies to any physical situation,
gives accurate results for $L\ll L_{\mathrm{nl}}$ and can
be generalized (higher-order perturbation) the take into account
multiple scattering events.
\section{Conclusion}
FPS in nonlinear waveguides is an important physical process that
enables photon-pair generation in well defined guided modes, with
high rate and low noise. Most of the experiments to date used the
scalar FPS process in which the pump photons and the scattered
photons have the same polarization. In birefringent waveguides,
vectorial FPS processes are also allowed: these vectorial scattering processes
involve photons with different polarizations. In this article, the
theory of FPS in nonlinear, birefringent, and dispersive fibers is
developed in the framework of the quantum theory of light. It gives
the general method for studying different vectorial FPS processes,
based on the first-order perturbation theory of scattering.

Different FPS occur in high- and low-birefringence fibers. These
cases have been studied separately. In the high-birefringence case,
photons with orthogonal polarizations can be generated. According to
the process used for their generation, they can be either entangled
in polarization or have their polarization correlated to their
wavelength. In the low-birefringence case, photon pairs with
polarization orthogonal to the pump field can be produced. This
generation methods has the advantage that polarization can be use to
separate the photons from the pump and that Raman noise is lower, as
has been proved recently \cite{Lin2006,Lin2007}. Our main concern
has been the spectral properties of the generated pairs. We showed
that very monochromatic photon pairs can be generated using the
vectorial FPS processes. Consequently, when picosecond pump pulses are used, 
it is possible to generated photon pairs with Fourier-transform limited spectra. 
We also noted that scalar and vector
processes may sometimes scatter photons into the same spectral
bands.

In order to apply the theory to practical design of photon-pair
sources, it may be desirable to extend it in different directions.
For instance, the theory can be extended to encompass more complex
dispersion relations, as required when working close to the
zero-dispersion wavelength or/and with photonic crystal fibers. This
can be done by modifying the dispersion operator in the quantum
nonlinear Schr\"odinger equations (\ref{QNLS}). Exotic dispersion
can dramatically modify the spectral properties of the generated
photons. Including the spontaneous Raman effect is important for
estimating the signal-to-noise ratio of a source. This can be done
by generalizing the nonlinear terms of the quantum nonlinear
Schr\"odinger equations (\ref{QNLS}) as explained in \cite{Lin2007}.
For deriving the quantum state of light at the output of the fiber
and the mean photons fluxes, the method explained in this paper
applies and can by followed step by step. For dealing with finite
pump duration (pump pulses), the easiest method is to use discrete
modes, as we did it this work. If the precise pump shape must be
taking into account, the theory has to be modified more seriously.
This can be done in the same way as in the scalar scattering case
\cite{Li2008a}, but that method is not rigorous. A better method for
dealing with pump pulses is to integrate (\ref{QNLS}) numerically,
as explained in \cite{Brainis2004}. 

Understanding the physics of FPS in birefringent media is important
for the present development of the field of quantum photonics. This
work contributes to this rapidly growing field by presenting the
principles underlying vectorial FPS and, as explained above, can be
extended in many different ways to match the need of particular
applications.

\begin{acknowledgments}
The author gratefully acknowledge support by the Philippe Wiener and
Maurice Anspach Foundation and the EU through the research and
training network EMALI (MRTN-CT-2006-035369).
\end{acknowledgments}


\end{document}